# Advanced Inorganic Halide Ceramic Scintillators

R. Hawrami[1], E. Ariesanti[2], H. Parkhe[2] and A. Burger[2]

[1]Xtallized Intelligence, Inc., Nashville, TN 37211
[2]Fisk University, Nashville, TN 37208

**Abstract**

Research in ceramic scintillators has steadily progressed alongside the research in bulk single crystal scintillator growth. As interest in faster scintillation material production with lower cost increases, more research on scintillating ceramics is needed. Research targeting optimization of optically transparent ceramics that can rival bulk-grown crystals grown may lower cost, increase yield, increase volume, and improve energy resolution in applications and systems currently using sodium iodide and alike. Ceramic scintillators that are dense (>5 g/cm$^3$), have high effective Z (>60), are bright (>40,000 photons/MeV), and are not sensitive to moisture as well as those that can be handled without protection are desired. Ultra-fast ceramic materials are also of interest. This paper presents an equipment design and technique to produce inorganic halide ceramic scintillators $Cs_2HfCl_6$ (CHC) and $Tl_2HfCl_6$ (THC). Improvements and optimization of CHC and THC ceramic scintillator fabrication are gauged by monitoring the energy resolution and peak position of $^{137}$Cs full energy peak at 662 keV. With a 1-inch diameter CHC ceramic scintillator, energy resolution of 5.4% (FWHM) and light yield of 20,700 ph/MeV are achieved, while with a 16-mm diameter THC ceramic scintillator, energy resolution of 5.1% (FWHM) and light yield of 27,800 ph/MeV are achieved. Decay times of 0.6 μs (21%) and 3.0 μs (79%) are measured for CHC and 0.3 μs (13%) and 1.0 μs (87%) for THC. Both ceramic CHC and THC scintillators have similarly good proportionality data when compared to their single crystal counterparts.

**Index Terms**

Ceramic scintillators. Cesium hafnium halides. Thallium hafnium halides. Inorganic halide crystals.

# I. INTRODUCTION

There has been a growing interest in employing scintillator material with better properties than NaI:Tl for radioisotope detection and identification. The ability to better identify different isotopes will require a scintillation material that has an excellent energy resolution, excellent proportionality, and high light yield.

Recent developments in cerium ($Ce^{3+}$) doped lanthanide halide single crystals, which include chlorides, bromides, and iodides, have produced inorganic scintillators that exhibit high light outputs, fast decay times, and outstanding energy resolutions, all excellent for various radiation detection applications [1, 2, 3]. However, producing these single crystals by bulk methods is expensive and they are difficult to grow in large sizes due to the anisotropic nature of these materials. For example, thermal expansion coefficients for the hexagonal $LaBr_3$ (space group: P63/m) along its c-axis and normal to the prismatic plane are $13.46 \times 10^{-6}/°C$ and $28.12 \times 10^{-6}/°C$, respectively. This difference in expansion coefficients can create large thermo mechanical stresses in the crystal during solidification process. Furthermore, these materials have extremely limited ductility and low fracture toughness in comparison to traditional halide salts. Cracks can be easily initiated and propagating along the crystal bulk. These factors limit the available crystal sizes, increase manufacturing costs, and hamper the widespread use of these materials for radiation detection applications [4].

Another recently published scintillator, $Cs_2HfCl_6$ (CHC), is an intrinsic, non-hygroscopic scintillator with many attractive physical and scintillation properties that are comparable or even better than NaI:Tl [5, 6, 7]. CHC was first reported as an example of non-hygroscopic compounds having the generic cubic crystal structure of $K_2PtCl_6$ [8]. The luminescence of intrinsic CHC and its homologue $Cs_2ZrCl_6$(CZC) were first studied in 1984 [9]. The first effort to grow intrinsic CHC was done with the Bridgman method [5]. Without dopant CHC's emission spectrum was centered around 400 nm, with a principal decay time of 4.4 μs, a light yield of up to 54,000 photons/MeV (when determined via comparison to a similar BGO crystal of known brightness, or 37,000 photons/MeV when compared to NaI:Tl), and energy resolution of 3.3% at 662 keV using a 0.65 cm$^3$ cubic sample [5].Pulse shape discrimination (PSD) between $^{137}$Cs gamma-rays and $^{241}$Am alphas using CHC resulted in a figure of merit (FOM) of 7.5 using the charge integration method, suggesting that it is possible to use CHC for particle discrimination [10].Following its re-discovery [5]subsequent papers has reported issues in growing CHC through the melt growth method [11, 12]. In a further study on the crystal growth and behavior of CHC and its variant, $Cs_2HfCl_4Br_2$ (CHCB), both CHC and CHCB crystals were prepared by melt compounding sublimed $HfCl_4$ with CsCl and CsBr to produce materials for Bridgman growth. Both crystals showed minimal moisture sensitivity [12]. Shown in the∅1-cm and ∅1-inch CHC crystals were an evidence of CsCl as a secondary phase, which was a result of a non-stoichiometric (CsCl-rich) melt composition due to the high vapor pressure of $HfCl_4$ during compounding. This secondary phase was verified using micro-X-ray fluorescence spectrometry [12]. The study also reported an evidence of a secondary phase in CHCB [12]. From a clear CHC sample light yield and energy resolution of 30,000 ph/MeV (when determined via comparison of a similar NaI:Tl crystal of known brightness) and 3.3%, respectively, with a primary decay component of 3.9μs, were measured. A sample of CHCB with a secondary phase present in the core had a light yield and energy resolution of 18,600 ph/MeV and 4.4%, and with a primary decay component of 2.0μs for CHCB. Further purification process of the starting materials indicated that only less than 60% of as-received $HfCl_4$ was pure [6]. Nevertheless, better purification of the growth precursors enabled the same researchers to grow clear and inclusion-free large diameter CHC and CHCB boules with 3.5% and 3.7% energy (FWHM at 662 keV) energy resolution, respectively [13]. Further improvements of this compound were done by replacing $Cs^+$ with $Tl^+$ [14]. This substitution increases both the density as well as the effective atomic number $Z_{eff}$, factors which directly determine photon detection efficiency of materials. 16mm diameter $Tl_2HfCl_6$ (THC) and $Tl_2ZrCl_6$ (TZC) crystals grown by Bridgman technique were reported [14]. Densities of 5.1 g/cm$^3$ and 4.5 g/cm$^3$ as well as effective atomic numbers ($Z_{eff}$) of 71 and 69 were measured for THC and TZC, respectively. Energy resolutions of 3.7% (FWHM) for THC and 3.4% for TZC (FWHM) at 662 keV were measured. Primary luminescence decay times of 1.1μs and 2.3μs for THC and TZC, respectively, as well as excellent proportionality for both materials, were observed [14].

The high vapor pressure of the growth pre-cursors makes growing CHC-type compounds by melt-growth methods challenging, let alone producing them with high growth yield. By taking advantage of the ceramic fabrication technique, the growth issues related with high vapor pressure and high melting point can be avoided. Moreover, lower cost, shorter growth time, higher growth or production yield, and more homogenous samples are attainable. In an initial research published on ceramic processing of inorganic halide compounds lanthanide halide and elpasolite families were chosen [4]. The elpasolite compounds were chosen specifically for their cubic structure (space group Fm-3m). Four high Z, lanthanum-based elpasolite halides, $Cs_2NaLaBr_6$, $Cs_2LiLaBr_6$, $Cs_2NaLaI_6$, and $Cs_2LiLaI_6$ were initially selected for the synthesis and characterization due to their interesting scintillation properties [4].

Due to all the downsides faced in producing high growth yield and well-performing bright scintillators, an alternative ceramic approach has recently been explored to address these manufacturability issues. In this ceramic approach, Xtallized Intelligence, Inc. (XI, Inc.) has chosen $K_2PtCl_6$ compounds $Cs_2HfCl_6$ (CHC) and $Tl_2HfCl_6$ (THC) with a cubic crystal structure in the Fm-3m space group, because the isotropic nature of the cubic structure of $K_2PtCl_6$ leads to minimal thermo mechanical stresses [8]. Additionally, CHC being non-hygroscopic is a good candidate material for a new ceramic scintillator. Compounds with cubic crystal structures, such as $Cs_2HfCl_6$ are highly desirable in forming optical ceramic scintillators. This paper will present the development of technique for successfully producing low cost, large diameter (up to one inch) CHC and THC ceramic scintillators with comparable scintillating performance as their single bulk-grown scintillator crystals.

## II. Experimental Methods

The experimental procedure for this project can be described in three steps. The first step involves the purification of the starting pre-cursors, followed by the synthesis of $A_2HfX_6$ (A= Cs or Tl, X= Cl) compounds. This initial purification of the starting pre-cursors is necessary as some pre-cursor materials were received with as-received impurities of 2N or lower. The second step is powder material preparation to produce different levels of powder or particle sizes (from 400 to 35 μm). The last step is ceramic fabrication by hot pressing using an automatic hydraulic press. The fabrication process started with small diameter ceramics (<1 inch) using the hotpressing method. Emphasis was placed on determining the ratio of grain sizes, temperatures, pressures, and chamber atmosphere, to produce $A_2HfX_6$ (A= Cs or Tl, X= Cl, Br, or I) ceramic scintillators.

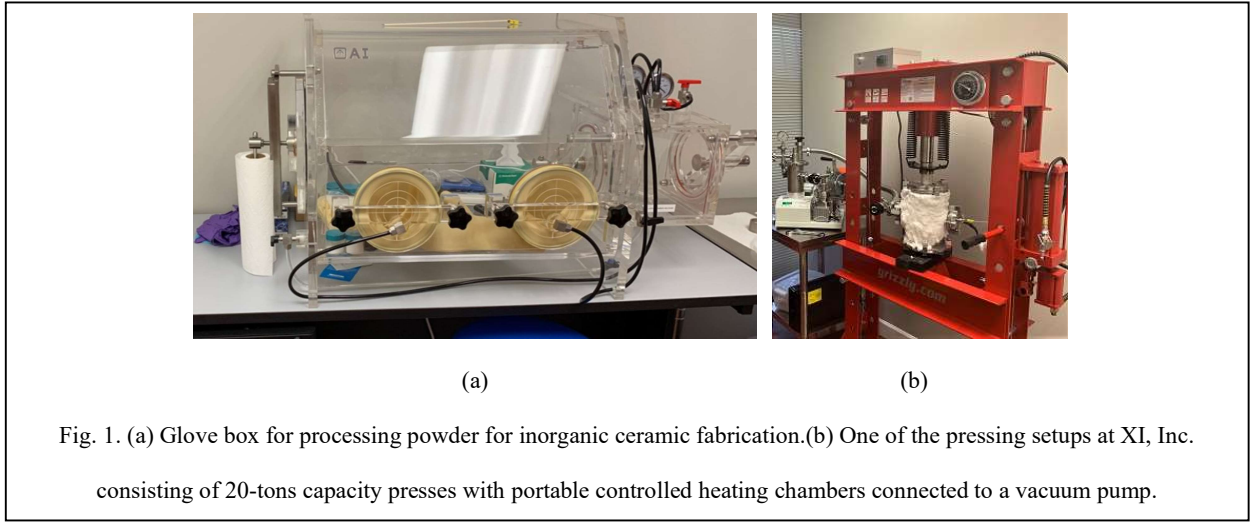

Fig. 1. (a) Glove box for processing powder for inorganic ceramic fabrication.(b) One of the pressing setups at XI, Inc. consisting of 20-tons capacity presses with portable controlled heating chambers connected to a vacuum pump.

Fig. 1(a) is a picture of a small glove box at XI, Inc. where the materials were processed and prepared before pressing. Processed powder will be prepared based on desired powder sizes. This preparation was accomplished by using a sieve-shaker to separate the powder to make and separate different powder sizes (400-35μm). Ceramic fabrication was done inside a portable chamber under high vacuum-and-heating for dehydration followed by pressing under a uniaxial hydraulic press (Fig.1(b)) for a for a period of six to twelve hours. Following this procedure and by continually improving available data for each compound, we managed to successfully produce large grain and uniform samples of $A_2HfX_6$ with performance close to their single bulk crystal counterparts. Characterization of the ceramic scintillators follows the same measurement procedure used to characterize single bulk-grown crystal scintillators and it has been described many times previously [6, 7, 13, 14]. Pulse height spectra were recorded with a standard gamma-ray spectroscopy system (i.e., NIM modules). A super bialkali PMT – R6231-100 was used for these experiments. A number of radioactive sources was used, including $^{137}$Cs, $^{22}$Na, $^{57}$Co, $^{133}$Ba and $^{241}$Am in order to cover a wide range of energies, ranging from 14 keV to 1275 keV. The resulting pulse height spectra were analyzed to yield photo peak information, peak position and energy resolution. Peak positions were used to estimate the light yield (using system calibration with a NaI:Tl crystal). Decay times were measured with a sample coupled to a PMT under irradiation from a radioactive source (e.g. $^{137}$Cs). The PMT's output was connected to a digital oscilloscope or a digitizer (CAEN DT5720C) and a number of waveform traces were recorded and averaged. The decay time constants were extracted using a multi-exponential fit.

## III. Results and Analysis

After material pufication and compound synthesis, a systematic study of ceramic fabrication runs was started, concentrating on three major dependent variables (particle powder size, heat level, and pressing level). The starting compound powders used were of various powder particle sizes starting from 400 to 35 μm particle sizes. These experiments have shown that the lower the paricle size was the better quality ceramics were produced. Fig. 2(a) shows the quality of fabricated CHC ceramic scintillators as a function of powder or particle size, while Fig. 2(b) shows the quality of fabricated CHC ceramic as a function of applied heat (i.e. temperature), and Fig. 2(c) shows the quality of fabricated CHC ceramic as a function of pressure amount.

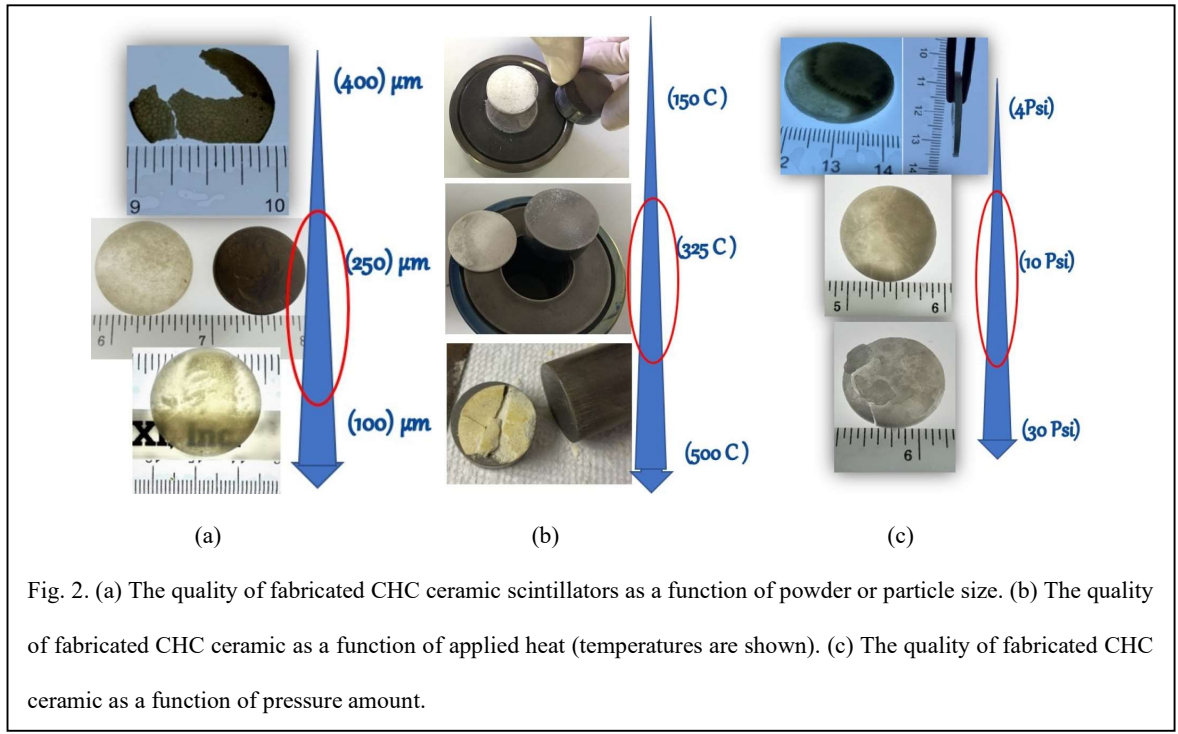

(a) (b) (c)

Fig. 2. (a) The quality of fabricated CHC ceramic scintillators as a function of powder or particle size. (b) The quality of fabricated CHC ceramic as a function of applied heat (temperatures are shown). (c) The quality of fabricated CHC ceramic as a function of pressure amount.

Fig. 3(a) shows a successfully fabricated, highquality1-in. diameter CHC ceramic scintillator. The thickness of the fabricated ceramic scintillator was approximately 2 mm, however, it was not uniform as indicated by parts of the ceramic that were translucent, and other parts that were more opaque. The more translucent part of the ceramic (indicated by a white circle in Fig. 3(a)) was extracted and used to collect a $^{137}$Cs spectrum shown in Fig. 3(b). Energy resolution of 5.4% (FWHM) at the full energy peak of 662 keV and light yield of 20,700 ph/MeV were measured. These results were close to the results commonly achieved by bulk-grown single crystals of CHC [6, 7].

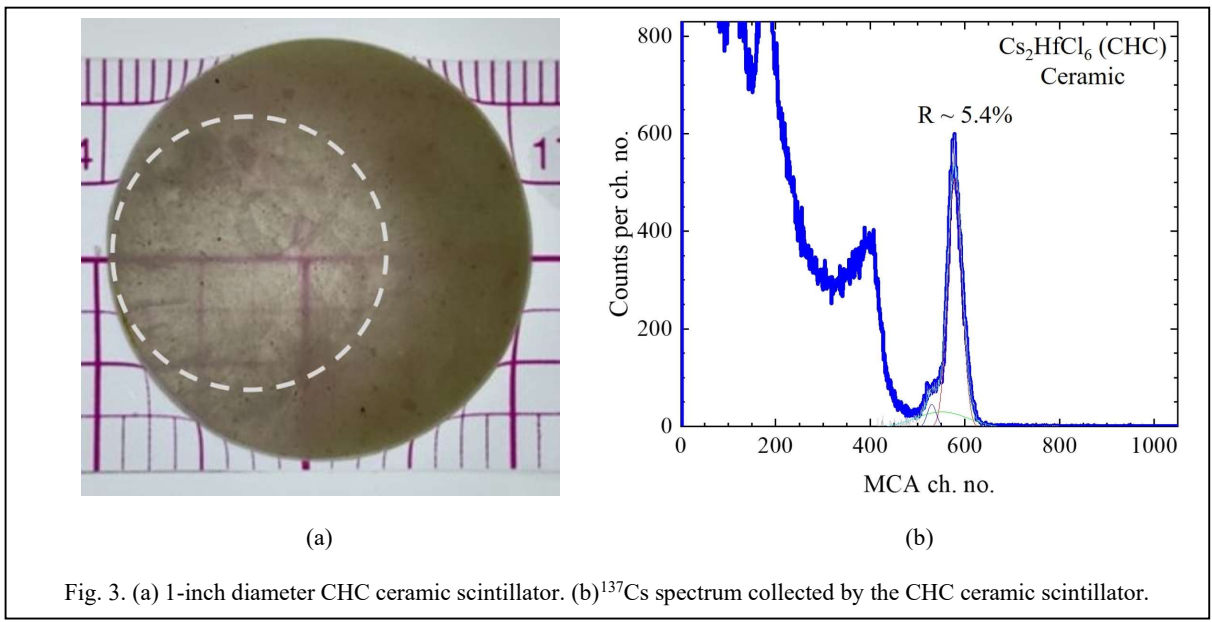

Fig. 3. (a) 1-inch diameter CHC ceramic scintillator. (b) $^{137}$Cs spectrum collected by the CHC ceramic scintillator.

Fig. 4(a) shows a successfully fabricated, high quality 16-mm diameter THC ceramic scintillator. The thickness of the fabricated ceramic scintillator was approximately 2 mm. The more translucent part of the ceramic (indicated by a blue ellipse in Fig. 4(a)) was extracted and used to collect a $^{137}$Cs spectrum shown in Fig. 4(b). Energy resolution of 5.1% (FWHM) at the full energy peak of 662 keV and light yield of 27,800 ph/MeV were measured. These results were close to the results commonly achieved by bulk-grown single crystals of THC, with the light yield result slightly better than the bulk-grown crystal [13, 14].

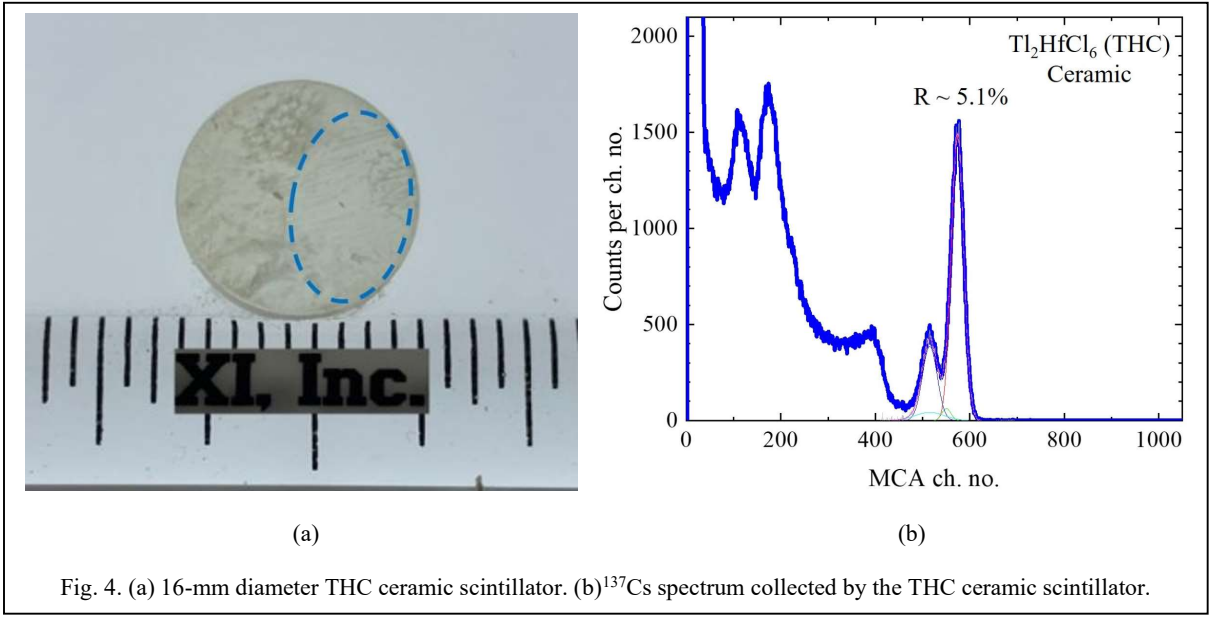

Fig. 4. (a) 16-mm diameter THC ceramic scintillator. (b)[137]Cs spectrum collected by the THC ceramic scintillator.

The time profiles for both CHC and THC ceramic scintillator are shown in Fig. 5. Decay times were calculated by fitting two exponential functions to either time profile, resulting in 0.6 μs (21%) and 3.0 μs (79%) for CHC and 0.3 μs (13%) and 1.0 μs (87%) for THC. Fig. 6(a) shows the time profile comparison between bulk and ceramic CHC. The primary decay time for the CHC ceramic scintillator is shorter than the bulk crystal, however, the fast decay time for CHC ceramic scintillator is longer than the bulk crystal. Comparison between the time profiles of bulk and ceramic THC shows that both decay components of the THC ceramic scintillator are faster than those of the bulk crystal counterpart (Fig. 6(b)).

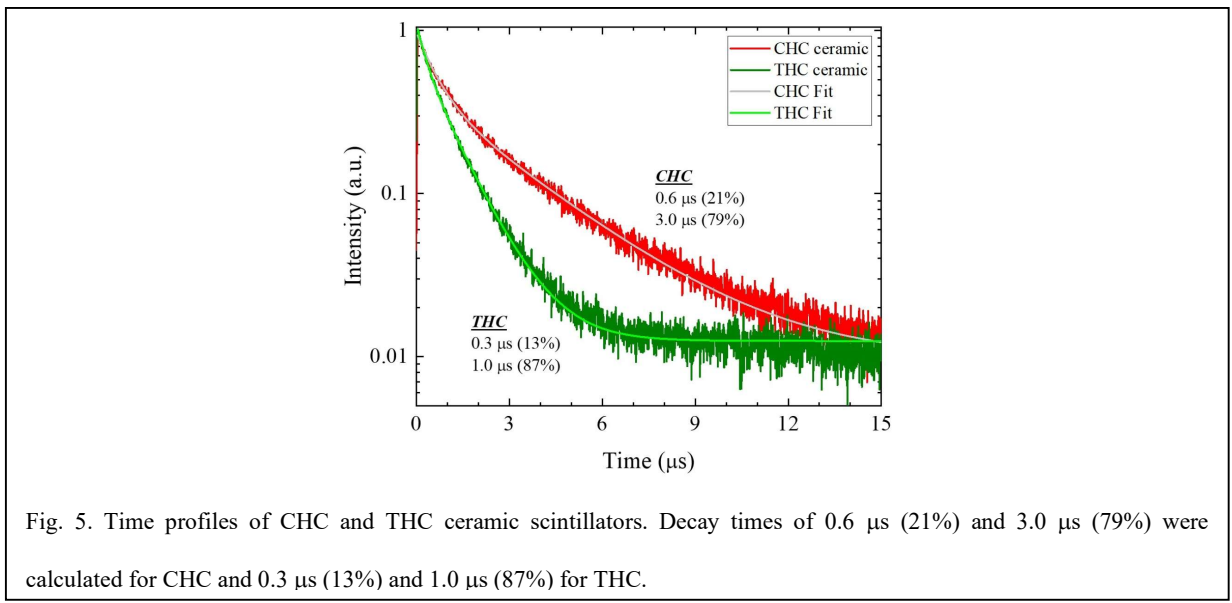

Fig. 5. Time profiles of CHC and THC ceramic scintillators. Decay times of 0.6 μs (21%) and 3.0 μs (79%) were calculated for CHC and 0.3 μs (13%) and 1.0 μs (87%) for THC.

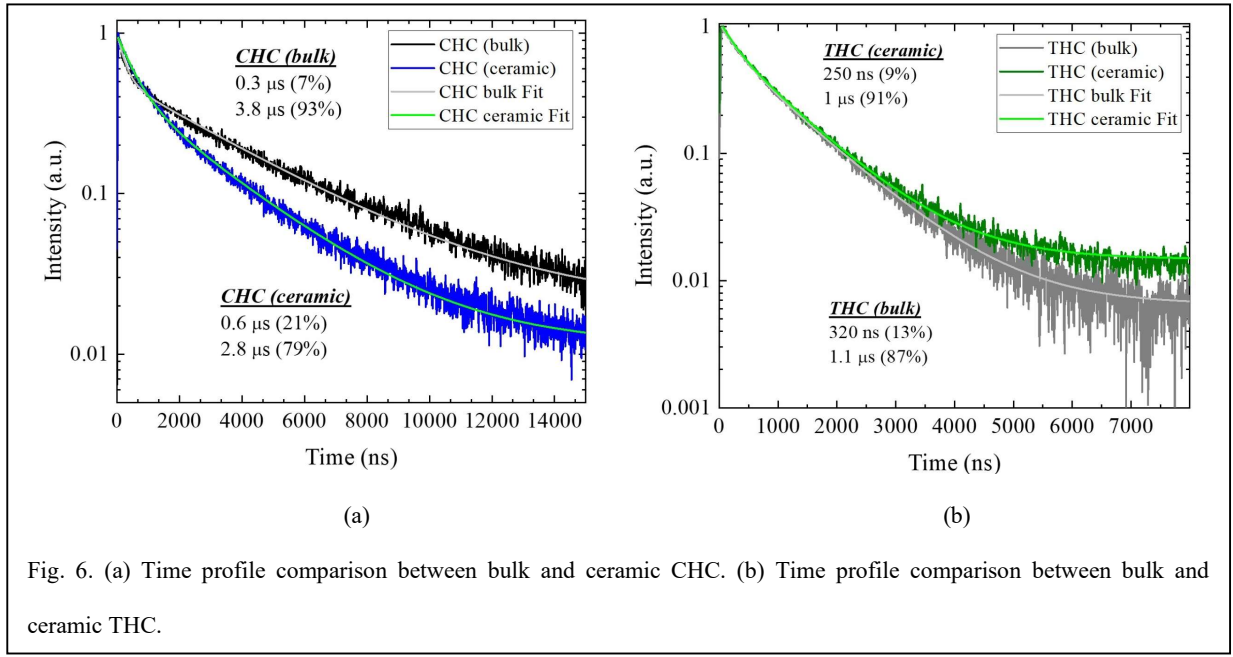

Fig. 6. (a) Time profile comparison between bulk and ceramic CHC. (b) Time profile comparison between bulk and ceramic THC.

Fig. 7(a) shows the non-proportionality data (relative light yield with respect to photon energy) comparison between single crystals NaI:Tl and BGO, as well as single crystal and ceramic CHC scintillators. Similarly Fig. 7(b) shows the non-proportionality data comparison between single crystals NaI:Tl and BGO, as well as single crystal and ceramic THC scintillators. Single crystal CHC is known for having an excellent proportionality (i.e., 0.95 < Relative Light Yield < 1.05) [5, 7, 6], shown in Fig. 7(a), and CHC ceramic scintillator also shows the same excellent proportionality. Single crystal THC has demonstrated good proportionality for photon energy beyond 60 keV [14, 13], shown in Fig. 7(b), and THC ceramic scintillator also demonstrates similarly good proportionality above 60 keV.

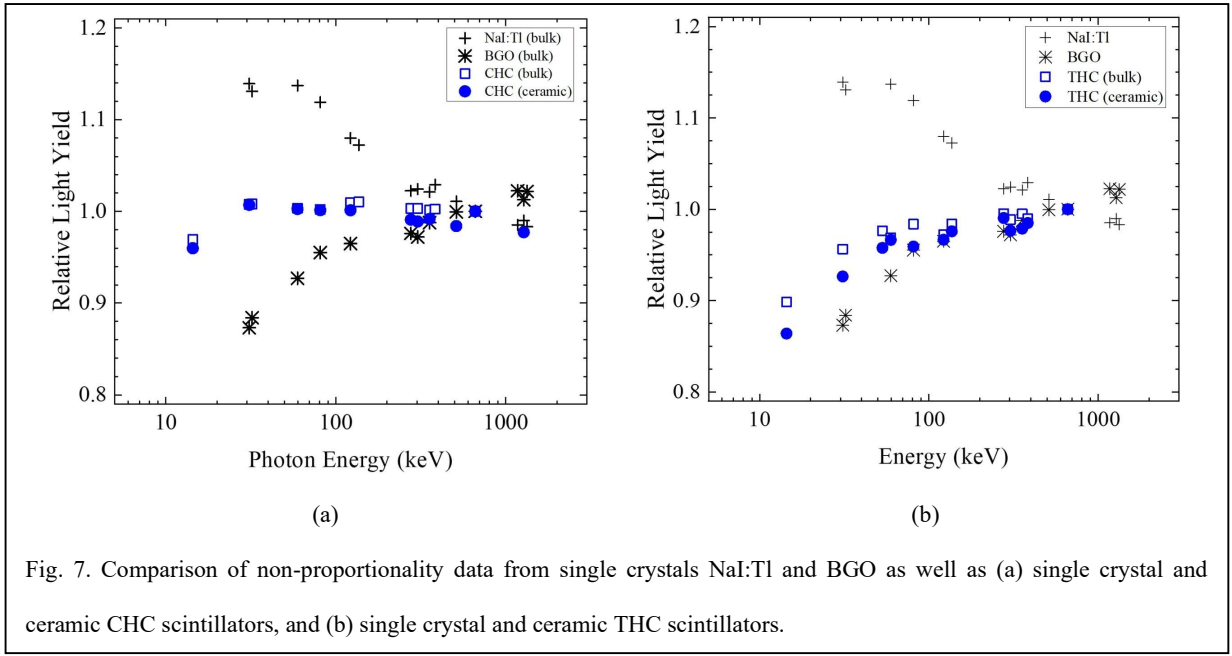

Fig. 7. Comparison of non-proportionality data from single crystals NaI:Tl and BGO as well as (a) single crystal and ceramic CHC scintillators, and (b) single crystal and ceramic THC scintillators.

Fig. 8 shows a comparison $^{133}$Ba spectra collected by 1″× 1″ NaI:Tl single crystal, 1 cm$^3$ BGO single crystal, 16-mm diameter × 3 mm CHC single crystal, and the CHC ceramic scintillator. For photon energies below 100 keV, for example for the characteristic x-ray peak at 31 keV and the gamma-ray peak at 81 keV, the spectra collected by the single crystal and ceramic CHC are similar. This similarity may be due to CHC attenuation length of 0.3 mm and 0.8 mm for photons at 31 keV and 81 keV, respectively. Therefore at thicknesses above 2 mm, most of the photons were attenuated. Gamma-ray peaks beyond 250 keV are more defined for the single crystal CHC due to counting statistics, however, better peak definition is expected for the ceramic CHC when longer counting time is employed. Overall, both single crystal and ceramic CHC have better peak definition due to its high energy resolution (when compared to NaI:Tl and BGO) and high light yield (when compared to BGO) as well as better proportionality.

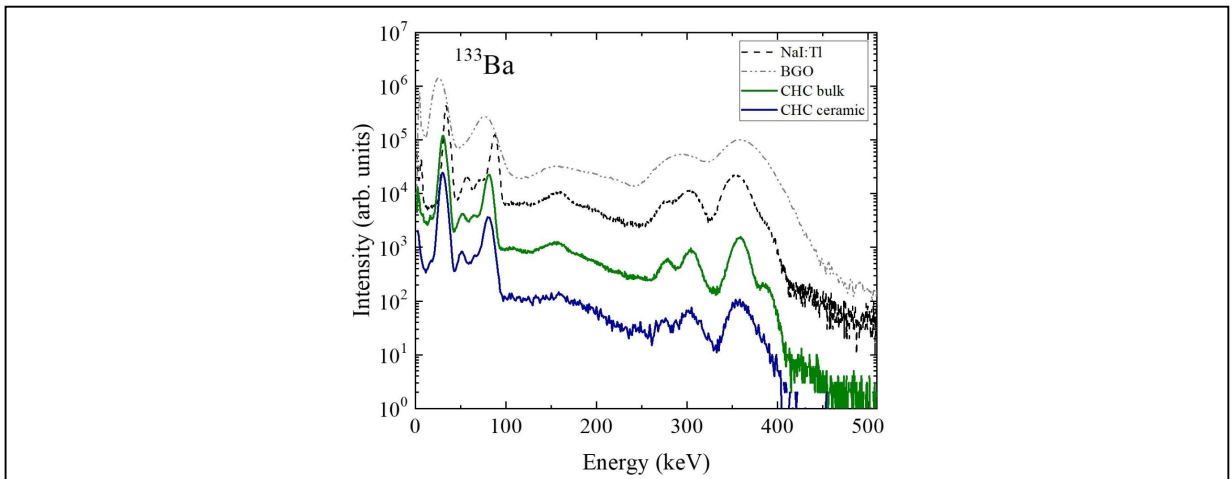

Fig. 8. Comparison of $^{133}$Ba spectra collected by 1″ × 1″ NaI:Tl single crystal, 1 cm$^3$ BGO single crystal, 16-mm diameter × 3 mm CHC single crystal, and the CHC ceramic scintillator.

Fig. 9(a) shows the improvement on the CHC ceramic scintillator performance over time, indicated by the run numbers (i.e., the smaller the run number the earlier in time the run was conducted). Energy resolutions (ER%) and the peak positions (Peak Ch. No.) for the full energy peak at 662 keV (the measurement settings were constant) were used as means to gauge the improvement of the CHC ceramic scintillator fabrication and performance. Fig. 9(b) shows the comparison of $^{137}$Cs spectra collected by selected CHC ceramic scintillators to show the improvement of the CHC ceramic scintillator fabrication and performance from the start to the current standing of the project.

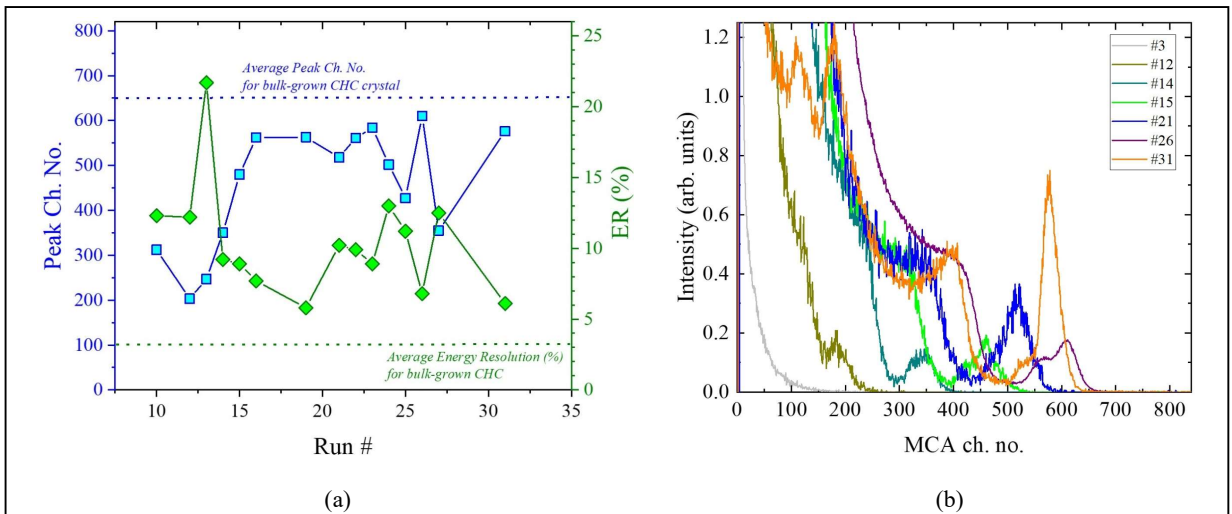

Fig. 9. Improvement of the CHC ceramic scintillator fabrication and performance from the start to the current standing of the project: (a) improvements in energy resolution of 662 keV peak and the peak position as a function of time (i.e., run number) and (b) comparison of $^{137}$Cs spectra collected by selected CHC ceramic scintillator fabrication and performance from the start to the current standing of the project.

## IV. CONCLUSIONS

The development of a technique to overcome issues related to scintillator crystal growth, such as cost, growth yield, and material waste, is described in this paper. The ceramic fabrication method is proposed to overcome these issues in order to produce well performing novel scintillators. melt crystals growth techniques facing on producing novel well performing scintillators. This paper presents an equipment design and technique to produce inorganic halide ceramic scintillators $Cs_2HfCl_6$ (CHC) and $Tl_2HfCl_6$ (THC). Improvements and optimization of CHC and THC ceramic scintillator fabrication were gauged by monitoring the energy resolution of $^{137}$Cs full energy peak at 662 keV and the peak position of the peak. With a 1-inch diameter CHC ceramic scintillator, energy resolution of 5.4% (FWHM) and light yield of 20,700 ph/MeV were achieved, while with a 16-mm diameter THC ceramic scintillator, energy resolution of 5.1% (FWHM) and light yield of 27,800 ph/MeV were achieved. Decay times of 0.6 μs (21%) and 3.0μs (79%) were measured for CHC and 0.3 μs (13%) and 1.0 μs (87%) for THC. The primary decay time for the CHC ceramic scintillator is shorter than the bulk crystal, however, the fast decay time for CHC ceramic scintillator is longer than the bulk crystal. Comparison between the time profiles of bulk and ceramic THC shows that both decay components of the THC ceramic scintillator are faster than those of the bulk crystal counterpart. Both ceramic CHC and THC scintillators have similarly good proportionality data when compared to their single crystal counterparts. Thin $A_2HfX_6$-type ceramic scintillators can be incorporated in spectroscopy applications to detect soft x-rays or low energy gamma-rays.

## V. ACKNOWLEDGMENTS

This work was supported by U.S. Department of Energy SBIR Grant #DE-SC0020816.